**Depicting urban boundaries from a mobility network of spatial interactions: A case study of Great Britain with geo-located Twitter data**

Junjun Yin[1, 2, 3], Aiman Soliman[1,2, 4], Dandong Yin[1, 2, 3] and Shaowen Wang[1, 2, 3, 4, *]

*1. CyberGIS Center for Advanced Digital and Spatial Studies*

*2. CyberInfrastructure and Geospatial Information Laboratory*

*3. Department of Geography and Geographic Information Science*

*4. National Center for Supercomputing Applications*

*University of Illinois at Urbana-Champaign, IL, 61801, USA*

\* To whom correspondence should be addressed. Email: shaowen@illinois.edu

**Abstract:** Existing urban boundaries are usually defined by government agencies for administrative, economic, and political purposes. However, it is not clear whether the boundaries truly reflect human interactions with urban space in intra- and inter-regional activities. Defining urban boundaries that consider socio-economic relationships and citizen commute patterns is important for many aspects of urban and regional planning. In this paper, we describe a method to delineate urban boundaries based upon human interactions with physical space inferred from social media. Specifically, we depicted the urban boundaries of Great Britain using a mobility network of Twitter user spatial interactions, which was inferred from over 69 million geo-located tweets. We define the non-administrative anthropographic boundaries in a hierarchical fashion based on different physical movement ranges of users derived from the collective mobility patterns of Twitter users in Great Britain. The results of strongly connected urban regions in the form of communities in the network space yield geographically cohesive, non-overlapping urban areas, which provide a clear delineation of the non-administrative anthropographic urban boundaries of Great Britain. The method was applied to both national (Great Britain) and municipal scales (the London metropolis). While our results corresponded well with the administrative boundaries, many unexpected and interesting boundaries were identified. Importantly, as the depicted urban boundaries exhibited a strong instance of spatial proximity, we employed a gravity model to understand the distance decay effects in shaping the delineated urban boundaries. The model explains how geographical distances found in the mobility patterns affect the interaction intensity among different non-administrative anthropographic urban areas, which provides new insights into human spatial interactions with urban space.

**Introduction**

Official urban boundaries are defined by government agencies for political and administrative purposes. Urban environments are conceptualized as spaces that are recreated and formed by human activities (Schliephake 2014). A fundamental question when using the administrative, "top-down", approach to defining urban boundaries is whether the outcome respects spatial interactions of humans. These interactions can take the form of trade, commerce, social connections, and political activities across borders. Urban boundaries that respect human interaction space are important to city planning, traffic management and resource allocation (Gao *et al.* 2014, Jiang and Miao 2015, Liu *et al.* 2015, Long *et al.* 2015). Many studies adopt a "bottom-up" approach to urban boundary delineation, where the geographical space is partitioned into small units and each unit is represented as a node within a network structure. A suitable community detection algorithm was applied to partition the network and associated geographical space based on the strength of human interactions between the nodes (Lancichinetti and Fortunato 2009). Different social and spatial human interactions were considered to establish the edges of the network connecting the nodes. For example, a large set of telephone call records were used to represent the network of human interactions across space to delineate urban boundaries in Great Britain (Ratti *et al.* 2010). Extending the previous method to different countries (Sobolevsky *et al.* 2013), the authors argued that this method yields cohesive geographical divisions following socio-economic boundaries. While other researchers use social ties of Twitter users to identify cohesive regions for different countries across the world (Kallus *et al.* 2015), they found evidence for dividing the urban space due to local conflicts and cross-country unifying trends that further support the "bottom-up" approach to mapping non-administrative anthropographic boundaries.

      A common finding from the aforementioned studies is that strongly connected urban regions in the form of communities in the network space yield geographically cohesive areas, despite different community detection methods and various forms of social and spatial human interactions were used. A general consensus is that those geographically cohesive areas are instances of the spatial proximity effects, where the

interaction strength between two urban regions decreases as the geographical distance between them increases (Fotheringham 1981). Spatial proximity is closely related to Tobler's First Law of Geography: "*everything is related to everything else, but near things are more related than distant things*" (Miller 2004). While it is intuitively logical, limited research has been carried out to quantitatively explain how the spatial interactions shape the forms of connected geographical areas (i.e., urban boundaries). One major reason is that while geographical distance may affect the interaction strength, it is not explicitly expressed in "virtual" human interactions, such as social ties or phone call initiation.

In this study, we describe a novel approach to delineating non-administrative anthropographic urban boundaries from a mobility network of spatial interactions. Specifically, the spatial interactions refer to actual movements of Twitter users, which were derived from more than 69 million geo-located tweets. Geo-located Twitter data are proven to be a useful source for studying human mobility patterns at large geographical scales (e.g. the national level) (Hawelka *et al.* 2014, Cao *et al.* 2015, Jurdak *et al.* 2015). Our approach provides a novel view of non-administrative units based on physical commutes rather than social ties or phone call initiation. A unique advantage is that non-administrative anthropographic urban boundaries can be delineated in a hierarchical fashion based upon different ranges of physical movement, which are inferred from mobility patterns of Twitter users.

We delineated the geography of urban boundaries in Great Britain by imposing a virtual fishnet over the islands of Great Britain. Twitter user movements were used to establish the connections between the fishnet cells to form a connectivity network, where each cell acts as a node within the network. We applied the map equation algorithm (De Domenico *et al.* 2015) to partition the network and associated geographic regions. We found that the collective mobility patterns of Twitter users in Great Britain are divided into several distance ranges ranging from short, intra- to inter-city movements with clear distinction points. The identification of connected regions at each of these distance ranges yielded hierarchical boundaries of urban spaces in Great Britain. As the depicted urban boundaries exhibited an evident instance of spatial proximity, we employed a gravity model to understand the distance decay effects in shaping the delineated urban boundaries. The model explains how geographical distances found in the mobility patterns affect the interaction strength among different non-administrative anthropographic urban areas. Our study connects human mobility research with the delineation of non-administrative anthropographic urban boundaries based on Twitter user spatial interactions, and provides new insights into human spatial interactions with urban geographical structures.

## 2. Background and Related Work

Urban regions are discrete components in a greater set of regions, with or without physical boundaries separating them (Jiang and Miao 2015). For political and administrative purposes, government agencies define various boundaries to partition urban space into spatial units at different scales, for instance: counties, census tracts and electoral districts. However, the spatial extents of these units often overlap and agglomerate depending how citizens perceive their activity space and interact with their urban environments (Lynch 1960). As connections are made between these units via various human activities, such as social-economic relations and commute patterns of citizens, certain units become more strongly connected than others. The boundaries of the agglomeration of these units are argued to reflect how people naturally interact with the geographical space, which is important for city planning (Hollenstein and Purves 2010), urban growth evaluations (Jiang and Miao 2015, Long *et al.* 2015), and traffic management (Gao *et al.* 2014).

Empirical studies have attempted to delineate such boundaries using a variety of methods and data. These methods can be generalized into two types: spatial clustering and network-based approaches. Spatial clustering approaches determine the boundaries based on the intensity of human activities anchored to geographic locations, for instance, locations of social media check-ins (Cranshaw *et al.* 2012, Sun *et al.* 2016), place descriptions from crowd-sourced Web content (Vasardani *et al.* 2013), and geo-tagged Flickr data (Stefanidis *et al.* 2013, Hu *et al.* 2015). While notable boundaries of urban areas were delineated, dynamic connections between different spatial units were not captured in spatial clustering approaches, where findings are discrete and independent areas reflect a high intensity of human activities.

On the other hand, network-based approaches delineate urban boundaries based on the intensity of human interactions between different spatial units, where each spatial unit is represented as a node and an edge is modeled by human interactions between two nodes. Such human interactions can take physical or virtual forms, such as trade, commerce, social connections, and political activity across geographical borders. For example, social ties of Twitter users were used to identify cohesive regions for different countries across the world (Kallus *et al.* 2015). Another example is a map of Great Britain redrawn based the communication network of phone call initiations (Ratti *et al.* 2010). In contrast, physical interactions form connections between nodes by individuals physically relocating from one node to another, which is referred to as a mobility network of spatial interactions in this study. Many studies have attempted to extract mobility networks from various data sources, such as census migration data (Rae 2009), GPS traces of moving vehicles (Rinzivillo *et al.* 2012) and taxi trip records (Liu *et al.* 2015), mobile phone call data (Sobolevsky et al., 2013; Zhong et al., 2014), social media check-ins (Liu *et al.* 2014), and geo-located Twitter data (Gao *et al.* 2014, Hawelka *et al.* 2014). These networks of human spatial interactions are then further explored to reveal clusters of interaction intensity, for example, using visual-analytics (Rae 2009) or community detection methods (Liu *et al.* 2015).

Researchers argued that those geographically cohesive areas taking the form of communities in the network space are related to the distance decay effect, which implies the interaction strength between two urban regions decreases as geographical distance between them increases (Liu *et al.* 2014). However, limited research explored the linkages between such effects and the characteristics of the underlying spatial interactions, which is critical for explaining how spatial interactions affect the shapes of connected geographical areas (i.e., urban boundaries). While geographical distance is not explicitly expressed in "virtual" human interactions, we argue that seeking insights from the mobility network of spatial interactions with the characteristics of underlying mobility patterns can help to explain how distance decay effects affect interaction intensity and the form of depicted urban boundaries.

## *2.1. Large-scale mobility network from geo-located Twitter data*

To construct a large-scale mobility network of human spatial interactions, capturing human movements with fine-grained spatial and temporal granularity is desirable. Low-resolution mobility data collected from census records are estimated and aggregated at census tract level and do not necessarily reflect movements of the same individuals (Rae 2009). To collect detailed mobility data of individuals, using GPS trackers tends to produce the most accurate records of individuals' movements, which means a high degree of recording accuracy of user locations and update frequency. However, the data often have limited spatial extent with a small group of people. For example, 182 and 226 volunteers participated in such mobility data collection in Zheng *et al.* (2008) and

Rhee *et al*. (2011), respectively. Other than tracking people directly, GPS data based on tracked vehicles (e.g. taxi) may only be accessible to a certain group of people (Kung et al., 2014). Another popular mobility data source found in literature is mobile phone call data in the form of Call Detail Records (CDR), where users' locations are estimated by cell tower triangulation with accuracy in the order of kilometers (González *et al.* 2008, Kung *et al.* 2014, Zhong *et al.* 2014). Such a dataset can cover a relatively large spatial extent (e.g., national level) (Sobolevsky *et al.* 2013) and a large population sample (Kung *et al.* 2014). However, due to privacy concerns, the mobile phone call data are not publicly accessible, which is not ideal for achieving scientific reproducibility.

On the other hand, it has become increasingly popular for researchers to exploit publicly accessible mobility data derived from Location-Based Social Media (LBSM) platforms (e.g., Foursquare and Twitter). However, there are limitations and challenges when directly extracting and using mobility data from LBSM data. For example, compared to GPS traces, the update frequency of an individual's location varies depending on when a user is posting a new geo-located tweet or check-in at a new place. LBSM data have also been criticized for lacking population representativeness, as not all people use social media or send geo-located messages (Kung *et al.* 2014). Another research challenge is to identify social media users, as a social media account is not equivalent to a real person in the physical world. A number of studies have looked into the demographic aspect of LBSM data, in particular, Twitter data (Steiger *et al.* 2015, Luo *et al.* 2016). While their methods are diverse, these studies suggest stricter criteria for filtering and extracting individuals' movements.

In this research, geo-located Twitter data are chosen as the source for constructing large-scale mobility networks of human spatial interactions and investigating detailed mobility patterns. A geo-located tweet is a Twitter message with an additional geo-tag expressed as a pair of geographical coordinates that represent the location of posting the tweet. Twitter provides a publicly accessible streaming API (http://dev.twitter.com/streaming) for open data access. The geo-located Twitter data present some unique advantages for our research. Specifically, the high-resolution location information enables researchers to identify multiple travel modes in user mobility patterns (Jurdak *et al.* 2015) while the large spatial coverage allows to compare mobility patterns (Hawelka *et al.* 2014) across many urban areas.

## 3. Materials and Methods

### *3.1. Geo-located Twitter Data and Data Processing*

For this study, data were collected using the Twitter Streaming API by specifying a geographical bounding box to retrieve all geo-located tweets that fall in the box. To ensure complete coverage of Great Britain, the bounding box covered the British Isles uses the lower left and upper right geographic coordinates (49.49, -14.85) and (61.18, 2.63) in the WGS84 datum. We implemented a data crawler to continuously collect data for seven months (June $1^{st}$ – December $31^{st}$, 2014) resulting in over 101.8 million tweets with a total data volume of 60 GB. The data crawler managed to download all the geo-located tweets for the given bounding box, as it did not encounter any issue of exceeding the data quota by the 1% policy mentioned by Hawelka *et al.* (2014). To showcase the overall spatial coverage of the collected geo-located tweets, the locations of all the collected tweets are shown in Fig. 1. The collective point-based visualization reveals the geography of cities (Leetaru *et al.* 2013). Notice the clusters with higher densities of tweets correspond to the locations of major cities.

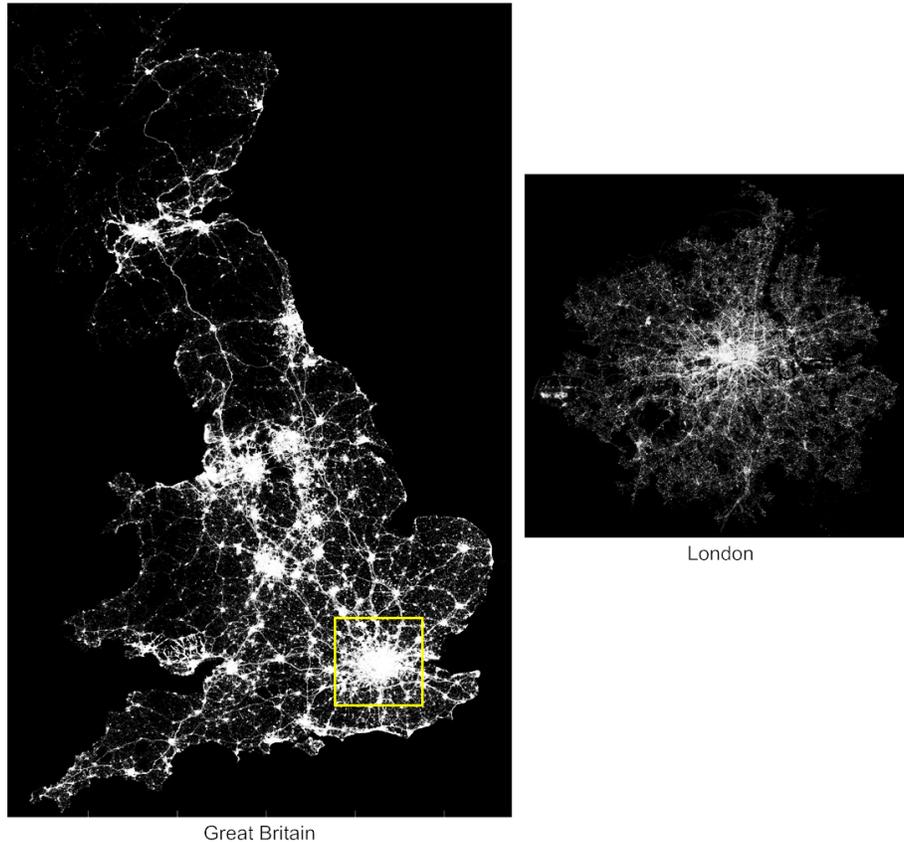

**Figure 1**. The spatial coverage of the collected geo-located Tweets for Great Britain (left) and London (right). Each point corresponds to an individual geo-located tweet. Note that Twitter activities are most active in urban areas.

The original location information embedded in every geo-tag includes both latitude and longitude coordinates. We examined "geo" attributes in each raw tweet and kept the one with location information derived from GPS recording rather than from geocoding. The points were projected into the British National Grid coordinate system to reduce the complexity in distance calculation. The geographical boundary of Great Britain was derived from the Office for National Statistics of UK (http://www.ons.gov.uk/ons) to further restrict the remaining tweets to be "domestic". Based on these restrictions, the filtered dataset contained 69,847,497 tweets made by 1,153,891 Twitter users. To remove tweets from non-human users, the raw tweets were further filtered using the following steps. First, duplicated messages were removed. We then removed non-human users based on unusual relocation by examining all of the consecutive locations of each user and excluded those with relocating speeds in excess of the threshold of 240 m/s, as suggested in Jurdak *et al.* (2015). Furthermore, to reflect the spatial interactions of residents rather than tourists, we imposed a condition that the time interval between a user's first and last recorded tweets should be more than 30 days. In other words, a user is identified as a resident only if he/she has stayed in the study region more than 30 days. The filtered dataset for our analysis contains 60,209,778 tweets from 824,712 Twitter users.

At this stage, each geo-located tweet is represented as a tuple $\langle user\_id, loc, t, m \rangle$, where $user\_id$ is an anonymous Twitter user's ID; $loc$ is the recorded location of the tweet as a coordinate pair; $t$ is the timestamp of the tweet; and $m$ is the actual content of the tweet. We constructed a trajectory for each Twitter user by appending all the recorded locations (with the same $user\_id$) in chronological order based on the timestamps. To protect Twitter users' privacy, the ID field was replaced with a randomly generated unique number and the message content was removed. The actual location of each geo-located tweet is only used for distance calculation and determining the geographical unit to which it belongs.

### 3.2. Mobility network of Twitter user spatial interactions

A Twitter user's movement is defined as the individual's geographical relocation or displacement (González *et al.* 2008). This is not equivalent to a "trip" taken by an individual, because, displacement includes situations where the time interval between two consecutive recorded locations is more than several days. To identify clusters of urban regional connectedness, Twitter users' movements are used to derive a connectivity network, where two urban regions connect when at least one user's movement begins in one and ends in another. These connections can be represented by an origin-destination (OD) matrix based on collective Twitter user displacements. This OD matrix is essentially a mathematical representation of a weighted directed graph $G \equiv \langle V, E_w \rangle$, where $V$ is a set of spatial nodes corresponding to the underlying urban regions, $E_w$ is a set of edges each representing the connection between a pair of nodes, and the weights are assigned by the accumulated volume of Twitter users' movements.

To build a spatial network at the national level, the basic units need to be determined to serve as spatial nodes of the connectivity network. Previous studies have suggested equidistant spatial tessellation to generate nodes, which uses Voronoi polygons for spatial partitioning (Rinzivillo *et al.* 2012, Zhong *et al.* 2014). One limitation of this tessellation approach is that it would decrease the spatial resolution of aggregated geo-located tweets, because Twitter users' location information is derived from embedded GPS within mobile devices. Another approach is to partition the study area into a grid of spatial pixels (Ratti *et al.* 2010, Liu *et al.* 2015). However, the grid size could potentially cause biases due to the Modifiable Area Unit Problem (MAUP) (Openshaw 1984, Wong 2009). To compare our investigation with the findings from similar studies, and avoid subjectively deciding the cell size, we performed statistical analysis of Twitter user mobility patterns in Great Britain and measured the distribution of collective Twitter user displacements and the radius of gyrations of individuals (González *et al.* 2008, Jurdak *et al.* 2015). The radius of gyration is a metric to distinguish mobility patterns of individuals, which is defined by Eq. (1):

$$r_g = \sqrt{\frac{1}{n}\sum_{i=1}^{n}(p_i - p_{centroid})^2}, where\ p_{centroid} = \frac{1}{n}\sum_{i=1}^{n} p_i \qquad (1)$$

Eq. (1) measures the accumulated distances of deviation from the centroid of all the recorded locations in an individual user's trajectory, where $p_i$ is one of the user's locations and $p_{centroid}$ is the center of mass of the user's trajectory. By examining the probability distribution of the radius of gyration, also known as the spatial dispersal kernel $P(r_g)$ (Brockmann *et al.* 2006), we chose 10 km as the cell size at the national level of Great Britain (Fig. 3 - c, with details provided in the next section). More importantly, as 10 km is the distinct geographic distance for separating two main groups of Twitter users in terms of the spatial coverage in Great Britain, a 10-km size cell serves as a mask for spatial partitioning. In this way, we focus on inter-connections among different urban regions with less attention to movements around a user's neighborhood (i.e., within 10 km radius), such as home or work places. Thus, we created a fishnet with 2784 cells. The cells of the fishnet act as proxies representing individuals' spatial coverage areas to focus on the inter-connectivity among cells and identify strongly connected cell clusters.

*3.3. Community structure of the network of spatial interactions*

Based on the derived mobility network of spatial interactions, which is a directed weighted graph, we further determined clusters of strongly connected spatial nodes, known as communities in the context of graph space. There are a variety of community detection algorithms that produce different results (Coscia *et al.* 2011). Our research adopts Infomap because it is suitable for treating directed weighted graph (Lancichinetti and Fortunato 2009).

Infomap identifies communities by minimizing the expected length of the trajectory of a random walker (Rosvall *et al.* 2010), which is shown below:

$$L(M) = qH(Q) + \sum_{i=1}^{m} p_i H(p_i) \qquad (2)$$

In Eq. (2), $qH(Q)$ is the entropy of movement among clusters and $\sum_{i=1}^{m} p_i H(p_i)$ is the entropy of movement within clusters. Specifically, $q$ is the probability that a random walker jumps from one cluster to another, while $p_i$ is the probability of movement within cluster $i$. This method can be intuitively tailored to describe strongly connected clusters of urban regions based on Twitter user movements. The detailed implementation of Infomap can be found online at http://mapequation.org. To apply the Infomap algorithm, the mobility network should be organized as a weighted and directed graph, on which we confirmed that an undirected graph cannot lead to a meaningful result (see Supplement Materials section 4).

*3.4. Distance decay effect and gravity model*

The clusters of urban regions in the form of communities often yield geographically cohesive urban areas. This phenomenon is likely related to the distance decay effects, where the interaction strength between two urban regions decreases as the geographical distance between them increases. A gravity model is often used to explain such effects, as is shown in Eq. (3), where $\langle T_{ij} \rangle$ and $d_{ij}$ denote the interaction from $i$ to $j$ and distance between two places, respectively, $k$ is a constant, and $P_i$ and $P_j$ are the population size of place $i$ and $j$, respectively. The distance decay function, $f(d_{ij})$, expresses the interaction strength decreasing with respect to the increasing geographic distance, where parameter $\beta$ reveals the distance impact on interaction strength. A greater $\beta$ indicates stronger decay where the interaction strength is more influenced by distance. While population size may not be an accurate indicator to describe the repulsion or attractiveness between places, the gravity model is usually fitted by using observed interaction strength and the distance between geographical entities (Liu *et al.* 2014).

$$\langle T_{ij} \rangle = k * \frac{P_i * P_j}{f(d_{ij})}, and\ f(d_{ij}) \sim d_{ij}^{\beta} \qquad (3)$$

In this research, the primary purpose for adopting the gravity model is not to find the best $\beta$ value to estimate the potential interaction strength among depicted urban areas. Interestingly, the distance decay effects are also found in human mobility patterns, which is attributed to the constraints of complex urban structure (Zhao *et al.* 2016). Therefore, we hypothesize that the distance decay effects affect the interaction strength of two geographic regions and ultimately depict the urban structures (e.g., urban boundaries) given that that urban structures are conceptualized spaces that are recreated and shaped by human activities (Schliephake 2014). The testing of this hypothesis likely reveals whether the depicted urban boundaries are random artifacts, and reflect how naturally people move across geographical regions.

## 4. Results

### *4.1. Collective Mobility Patterns of Twitter Users in Great Britain*

We modeled different aspects of mobility patterns of Twitter users. These patterns include: the number of visited locations per user, the collective user displacements, and the radius of gyration of individuals to identify distinct distance ranges. We utilized these distance ranges to partition the geographical space of Great Britain into fine-grained cells and established the connectivity among these cells to delineate non-administrative anthropographic urban boundaries.

We found that the cumulative distribution function of the number of locations visited by each Twitter user follows a two-tier power law distribution (Fig. 2). Majority of the data follow a truncated power-law distribution $P(X \geq x) \sim x^{-\alpha} e^{-\lambda x}$, where $\alpha = 1.24, \lambda = 0.00132$; and the tail part (less than 2% of the entire population) follows a power-law distribution $P(X \geq x) \sim x^{-\alpha}$ with $\alpha$ value 3.2. The distribution was found to be consistent over each month (i.e., June to December, 2014), which has a slight offset in the truncated power-law distribution (the mean α value is 1.26 ± with a 0.05 $\sigma$ and the mean $\lambda$ value as 0.00134 ± with a 0.0002 $\sigma$).

The two-tier power law distribution indicates that the collective behavior of Twitter users visiting different locations can be well approximated with a (truncated) Lévy Walk (a random walk) model (Rhee *et al.* 2011, Reynolds 2012), which has also been identified in many human mobility studies using different mobility data (Zhao *et al.* 2016). The similarity among the distributions suggests that the mobility data extracted from geo-located tweets are temporally stable, at least at monthly intervals, which indicates that our approach using Twitter user mobility to delineate urban boundaries is viable. In addition, the Lévy Walk model reveals the diversity regarding the number of visited locations per user, which suggests a level of "randomness" in Twitter user movement across space. It, in turn, justifies our choice of using the map equation as the community detection algorithm to identify the clusters of urban regional connectedness using large-scale Twitter user movement data.

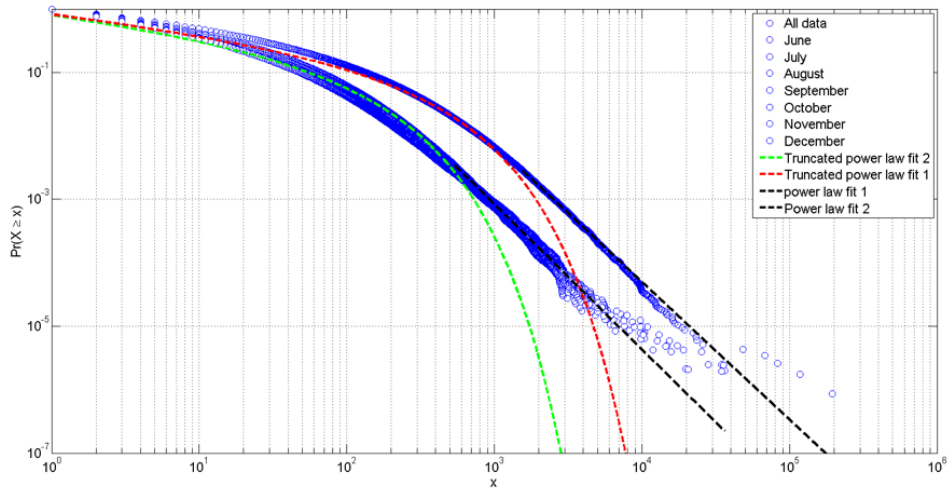

**Figure 2. Cumulative distribution of the number of locations visited by each Twitter user during different timespans**

We then studied two aspects of the Twitter user mobility patterns: the distribution of Twitter user displacement and the radius of gyration. Twitter user displacement refers to the distance between two consecutive locations in a user's trajectory using a straight-line distance metric. The radius of gyration describes the deviation of distance from the center of mass of a user's trajectory. The probability distributions of the collective user displacement $P(d)$ and radius of gyration $P(r_g)$ are presented in Fig. 3, where the fitting method for identifying different distance ranges is derived from Jurdak *et al.* (2015). The probability distribution of the collective displacements can be approximated by $P(d) \sim \lambda_1 e^{-\lambda_1(d-d_{min})}, d_{min} = 10\ m$ from [10 m, 70 m] (accounting for 3% of the population), $P(d) \sim \beta\lambda_1 d^{\beta-1} e^{-\lambda_1(d^\beta - d_{min}^\beta)}, d_{min} = 100\ m$ from [100 m, 70 km] (93% of the population), and $P(d) \sim d^{-\alpha}$ [> 70 km] (4% of the population). The displacement distance between 70 m and 100 km can be further approximated by two power law distributions with a cut-off point at 4 km (55% distances are less than 4 km and 40% distances between 4 km and 100 km), which indicates the urban movement captured by the geo-located tweets reveal two different modes: inter-city and intra-city movement. In short, these fitting functions suggest the existence of multi-scale or multi-modal urban movements captured from Twitter users in Great Britain, which means the geographically cohesive, non-overlapping urban areas identified in the next section are not just a result of short distance movements but emerge naturally from the broader Twitter user mobility patterns. Note that a similar multiphase pattern was observed in Twitter user displacements in Australia, but with slightly different distance ranges (Jurdak *et al.* 2015).

Further, we analyzed the distribution of radius of gyration $P(r_g)$ to understand the collective movements of individual Twitter users rather than separate displacements. The $P(r_g)$ of Twitter users in Great Britain can be approximated through a combination of three functions: $P(r_g) \sim \lambda_2 e^{-\lambda_2(r_g - r_{g_{min}})}$, $r_{g_{min}} = 10\ m$ from [10 m, 30 m], $P(r_g) \sim \beta \lambda_2 r_g^{\beta-1} e^{-\lambda_2(r_g^\beta - r_{g_{min}}^\beta)}$ from [50 m, 10 km], and $P(r_g) \sim r_g^{-\alpha}$ from [10 km, 100 km], where these three functions account for 92% of all the users. This suggests that there are three primary types of users that: (1) tend to stay at one location or at nearby locations when they tweet, or (2) tend to move at the intra-city scale when they tweet, or (3) tend to exhibit a large spatial coverage. (1) and (2) account for approximately 53% of all users. Note that the accuracy of these values for defining the distance bounds depends upon location accuracy of each geo-located tweet. These findings are consistent with the ones in the literature on human mobility, where the radius of gyration of human movement is bounded to different distance ranges (Brockmann *et al.* 2006, González *et al.* 2008). Interestingly, the $P(r_g)$ over the greater London region can be fitted by similar functions. However, as it reflects intra-city level mobility patterns, there is no distinct distance range to indicate the large spatial coverage. The distance decay effects found in both user displacements and the radius of gyration shows evidence of spatial proximity in Twitter user movement. It explains that the communities of urban regions within the graph space are geographically close but can be separated from other groups, which results in the delineation of urban boundaries based on the collective spatial interactions of Twitter users.

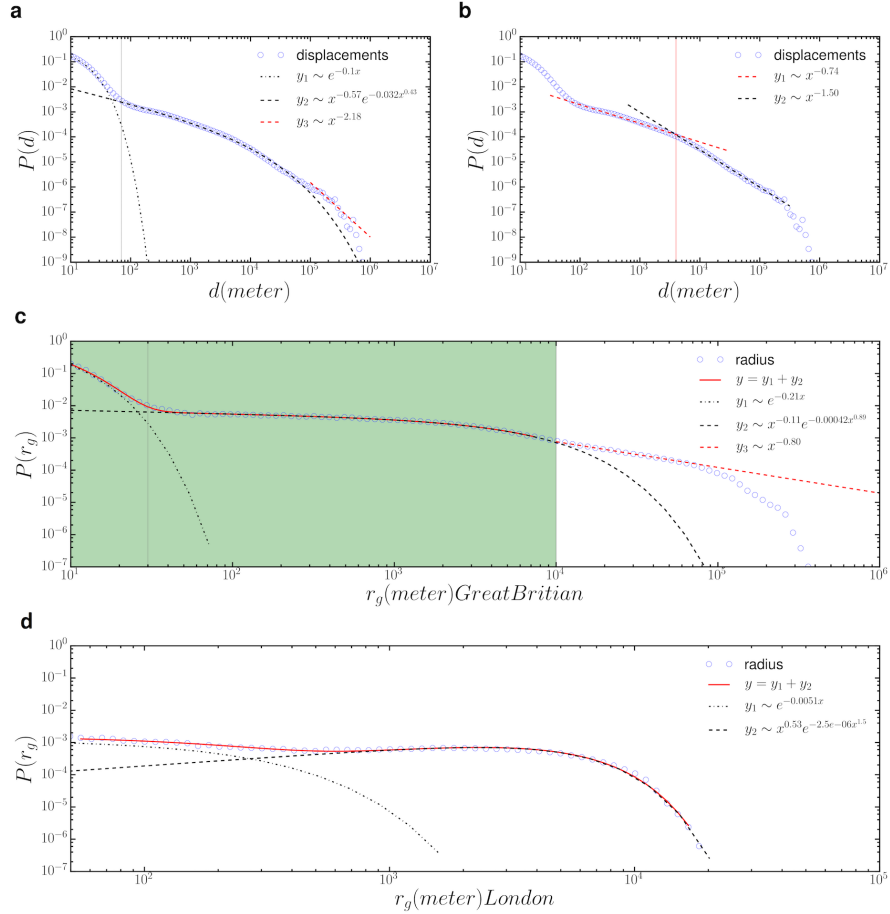

**Figure 3. The probability distribution of Twitter user displacements and radius of gyration:** (a) $P(d)$ is approximated by an exponential, a stretched-exponential and a power-law function, (b) the distance between [70 m, 70 km] is approximated by a double power-law function, (c) $P(r_g)$ of Great Britain is approximated by the combination of an exponential, a stretched- exponential and a power-law function, and (d) $P(r_g)$ of London is approximated by an exponential and a stretched-exponential function. The green patch shows the distance range.

## *4.2. Redrawing Great Britain's Urban Boundaries*

The mobility network of Twitter users' spatial interactions was constructed by nodes representing 10 km by 10 km fishnet cells, where 10 km is the distinct geographic distance for separating two main groups of Twitter users in terms of the spatial coverage (i.e., radius of gyration) in Great Britain (see Fig. 3(c)). The fishnet cells act as proxies representing individuals' spatial coverage areas to focus more on the inter-connectivity among cells and identify cell clusters. It provides an adequate resolution for a country-wide investigation (Ratti *et al.* 2010). The edges of this network were derived from the number of directed Twitter user displacements between each pair of cells. Coherent geographic regions were identified as individual fishnet cells showing more internal user movements compared to user movements across the cell boundaries to neighboring cells. To help readers who are not familiar with the geographic context in Great Britain better interpret the derived boundaries, two additional layers (i.e., locations of airport fields and population-weighted-centroids of workplace zones in Great Britain) are added in the background of the figures.

Fig. 4 shows the delineated urban boundaries based on Twitter user displacement distance less than 4 km, greater than 4 km, greater than 10 km, and using all available displacements together compared to the administrative boundaries of Great Britain. One clear observation in both the coarse and fine delineations is that most of the geographic divisions are centered on large urban cores with relatively dense populations. These results are expected given that most of the tweets originated from urban centers. However, what is remarkable is the performance of our approach to dividing the remaining space between cities. We found that restricting the trip distance results in different delineations of the catchment area around these centers. For example, one could explain these effects as a manifestation of the underlying gravity law (Simini *et al.* 2012) and the distance decay effect on attracting movers (González *et al.* 2008). Specifically, our approach performed well in terms of dividing the entire space with minimum gaps. Empty cells were found in regions where no or few Twitter users had visited (e.g. forests, agriculture) especially when restricting the analysis to short-distance Twitter user displacements.

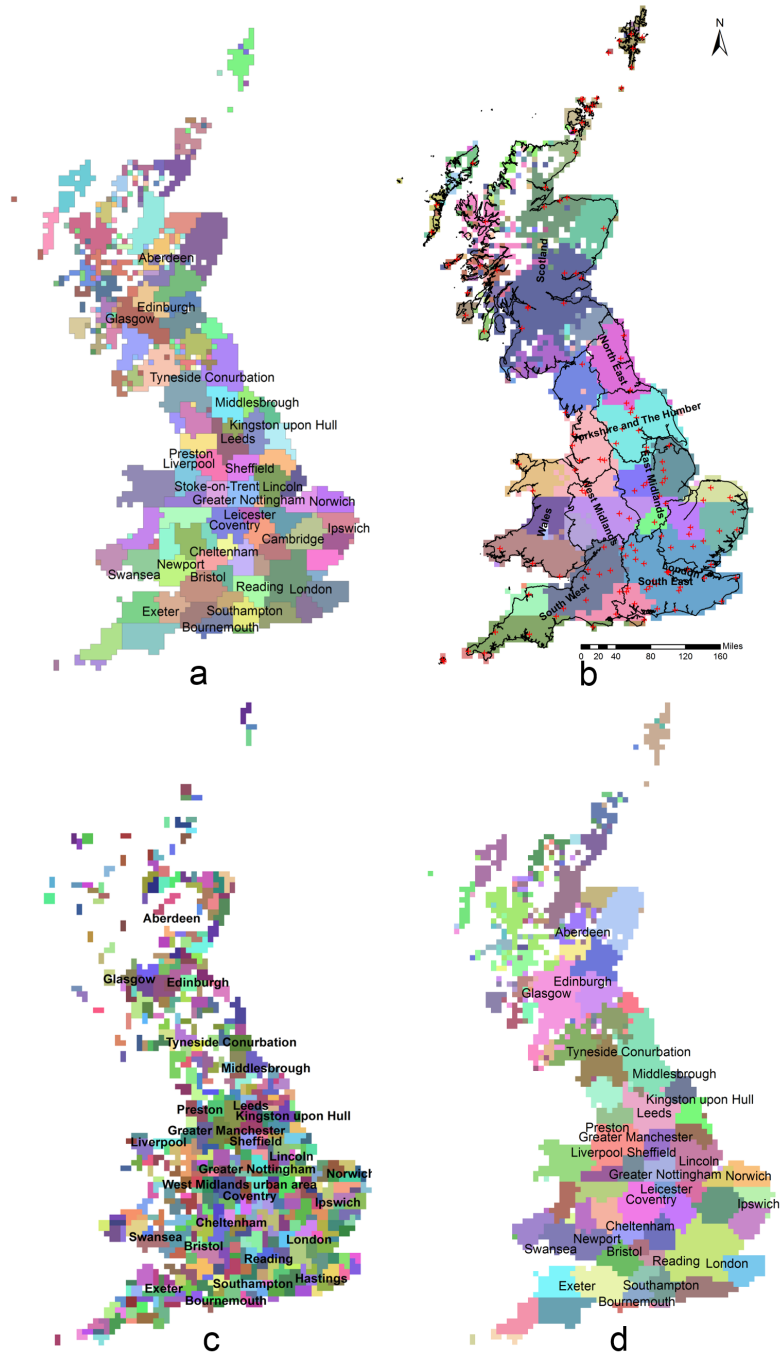

**Figure 4. The community structure from collective Twitter user displacements reveals non-administrative anthropographic urban boundaries:** (a) all displacements with L(M) = 7.8, (b) displacements longer than 10 km (L(M) = 8.5), red symbols are the locations of airport fields in Great Britain, (c) displacements shorter than 4 km (L(M) = 4.5), and(d) and displacements longer than 4 km (L(M) = 8.1). Each color represents a unique community with more Twitter user displacements among the cells compared to others. Major cities (urban audit functional areas) and NUTS (Nomenclature of Territorial Units for Statistics - 1) are displayed as labels.

Regional boundaries inferred from short distance Twitter user displacements (less than 4 km) exhibit very small and fragmented regions, which is probably related to daily commuting around a user's home location. Redrawing the boundaries based on longer distance displacements produces more cohesive, large regions. For example, by partitioning the space based on displacements greater than 10 km created regions that are comparable to the NUTS regions (Fig. 4 - b). However, the power of this novel mapping technique is not to reproduce the partitions already known, rather it is to point out unconventional boundaries. For example, the boundaries between England and Wales were found to be more diffusive compared to the abrupt boundary of England and Scotland. Moreover, the city of London has a wider visitor catchment area that extends beyond the authoritative boundaries of the city. Increasing the displacement distance results in revealing the large region connected to London (Fig. 5).

The patterns revealed from Twitter users' mobility are comparable to the patterns inferred from the network of landline phone calls (Ratti *et al.* 2010). For example, the region of Wales appears to consist of three communities as found in the connectivity of both phone calls and long distance movements. However, the regions extracted from the mobility network seem to be more spatially consistent with minimal spatial gaps compared to the partitions extracted from the network of landline phone calls.

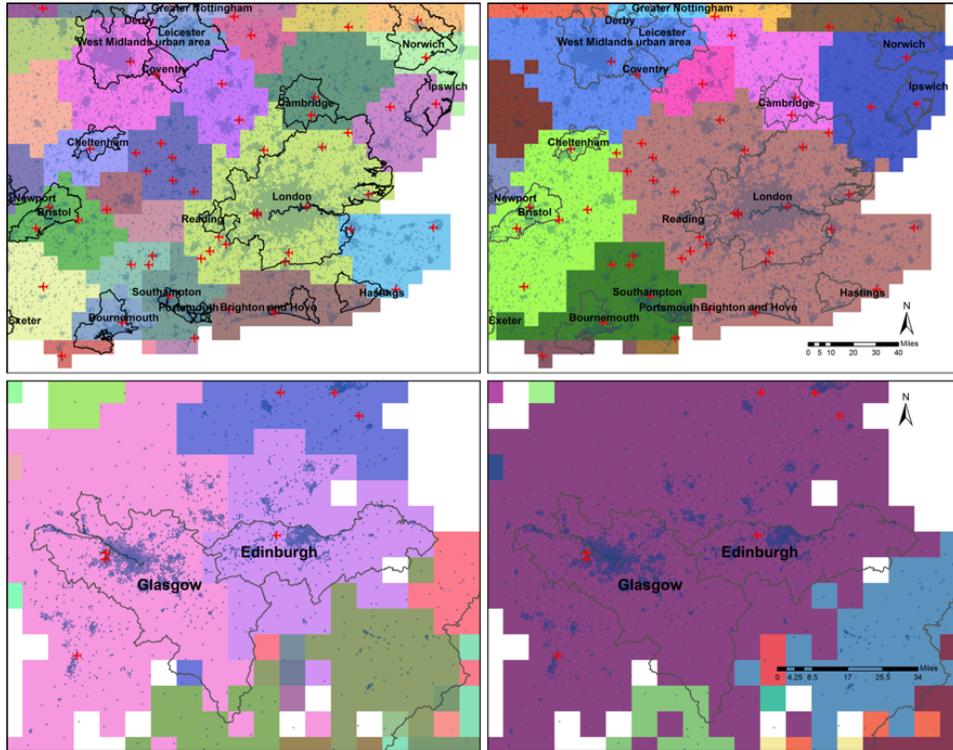

**Figure 5. The non-administrative anthropographic regions inferred from Twitter user displacements greater than 4 km (left) and 10 km (right) in comparison with major cities in England (upper figures) and Scotland (bottom figures)**. Each color represents a unique community. Including short distance movements has increased the power to differentiate the influence of nearby cities such as Glasgow and Edinburgh (lower left), while restricting the analysis to longer distance movements grouped travelers from the two previous cities into the same community (lower right). Red symbols are the locations of airport fields and gray points are the population-weighted-centroids of workplace zones in Great Britain.

A detailed study was conducted over the greater London region to reveal the intra-city spatial interaction patterns. Since the captured Twitter user movements were on an intra-city level (in comparison to the national level in Fig. 3-c), there was no distinct distance range to separate Twitter user spatial coverage in terms of radius of gyration (see Fig. 3-d). We chose 1-km cell size as suggested by Liu *et al.* (2015).

The spatial partitions that derived from a fine grid of 1-km cells and used all available Twitter user trips without any restriction on trip distances yield geographical boundaries comparable to some of London's boroughs (Fig. 6). However, some areas are shown to be more cohesive and display greater spatial interactions across the administrative boundaries, for instance, central London than others. Although, these results suggest that Twitter users seem to be localized over certain areas of the city most of the time, some regions do exhibit long distance interaction patterns. For example, the separate geographical areas in the south of Hillingdon which includes Heathrow Airport exhibits more connectivity to central London than its surrounding areas, which is explained by the usual flight passenger routes. The technique also reveals some of the emerging communities around the borders due to the spatial interminglement of both communities. For example, East Barnet and West Enfield seem to have higher interactions than those in the emerging cohesive zone between the two boroughs.

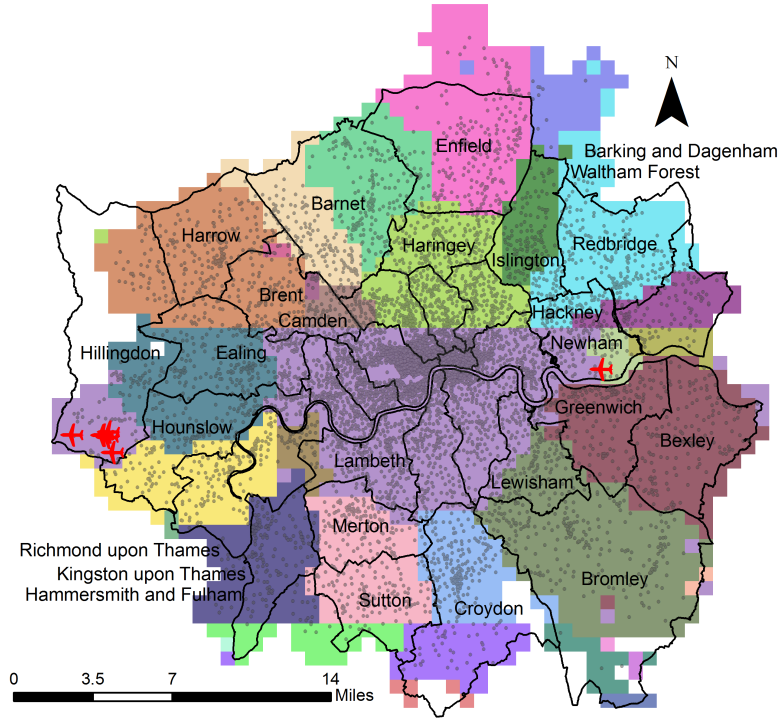

**Figure 6. Non-administrative anthropographic boundaries inferred from collective Twitter user displacements in the city of London compared to the boundaries of London boroughs (L(M) = 8.1)**. A fine fishnet of 1 km cells was used to identify the detailed connectivity patterns based on all of the Twitter user displacements in the area. Each unique color represents a different non-administrative anthropographic region. Notice that some remote regions like the airport (light purple region in south Hillingdon) share the same class with downtown because it is well connected despite the geographic separation. Red symbols are the locations of airport fields and gray points are the population-weighted-centroids of workplace zones in London.

In this study, we imposed a virtual fishnet to partition the geographical space over Great Britain. Alternatively, we had used the ward divisions as spatial units for aggregating Twitter user movements, which is the finest administrative boundaries of Great Britain (see Supplement Materials section 3). The derived communities in the network space are similar to the ones from using fishnet approach. The strongly connected communities also yield geographically cohesive, non-overlapping urban areas. However, as the ward division is still defined administratively purpose, the polygonal units tend to be geographically continuous. It causes problems to aggregate regions that do not have Twitter coverage into certain clusters. Aggregating Twitter user movements at the ward level also imposes more apparent concerns of the mismatch of the overall population, where less populated areas were overly represented and connected into large areas.

We should be aware that using different fishnet cell-size to partition the space will produce different mobility networks and can potentially lead to different delineations of the urban boundaries. While 10 km cell-size fishnet was applied at national level in this study, we also carried out an experiment by arbitrarily setting the cell-size to 5 km at the national level (see Supplement Materials section 4). Note that the cell-size could be set to any value, such as 4.9 km or 5.1 km. The fishnet with smaller cell-size (i.e., 5 km) produced more and smaller strongly-connected communities within the network space. Hence, the spatial resolution of the fishnet cells does affect the outcome from the community detection method employed in this study, where fishnet with smaller cell-size leads to more discrete and locally connected (i.e., smaller) clusters of urban areas. Such an effect can be explained by the probability distributions of the radius of gyrations of individual Twitter users. The probability of distance that deviates from a user's center location decays with a stretched-exponential function from [50 m, 10 km], which means the movements from Twitter users with smaller spatial coverage dominate the delineation of the connected urban areas. To avoid arbitrarily deciding the cell size, we studied the probability distributions of the radius of gyrations of individual Twitter users and selected 10-km as the cell-size, which is the distinct geographic distance for separating two main groups of Twitter users in terms of their spatial coverage at the national level. In addition, this choice enabled us to focus on the inter-connections among different urban regions with less attention to movements around a user's neighborhood (i.e., within 10 km radius), such as home or work places.

*4.3. Explaining the distance decay effect with a gravity model*

As the depicted urban boundaries exhibit a strong instance of spatial proximity, a gravity model is employed (Eq. (3)) to explain how distance decay effects found in the mobility patterns affect the interaction strength between the derived non-administrative anthropographic urban areas. In this model, the distance between two derived urban areas is measured by the geodetic distance between the centroid locations of the two. $P_i$ and $P_j$ are the observed interaction between urban area $i$ and $j$, which are measured by the aggregation of movement flows in each area. In particular, we set the distance decay parameter $\beta$ value as 0.8: (1) As we hypothesize that the distance decay parameters found in the underlying mobility patterns potentially contribute to $\beta$ in the gravity model (2) and we chose the 10-km cell size based on the collective Twitter user mobility pattern regarding radius of gyration, where the distance decay parameter is 0.8 when radius of gyration $r_g > 10\ km$ (Fig. 3(c)). We found that the gravity model indicates a strong linear correlation between the observed versus the estimated interaction strength with $R^2 = 0.89$ and $p - value < 0.01$. This confirms that the depicted urban areas are instances of spatial proximity effects, where the strength of human (in this case, Twitter user) spatial interactions between two urban regions decreases as the geographic distance between them increases.

The well-fitted gravity model provides support that the depicted urban areas are not just random artifacts but reflect how people naturally move across geographic regions. More importantly, since we have used a mobility network to delineate the boundaries, the distance decay effects are well related and explained by the distance decay parameters found in the underlying mobility patterns. To elaborate, the spatial interaction strength decreases along with the decay for the probability of longer distance Twitter user movements, and eventually stops at certain spatial extent, which leads to more geographically cohesive cluster of urban regions.

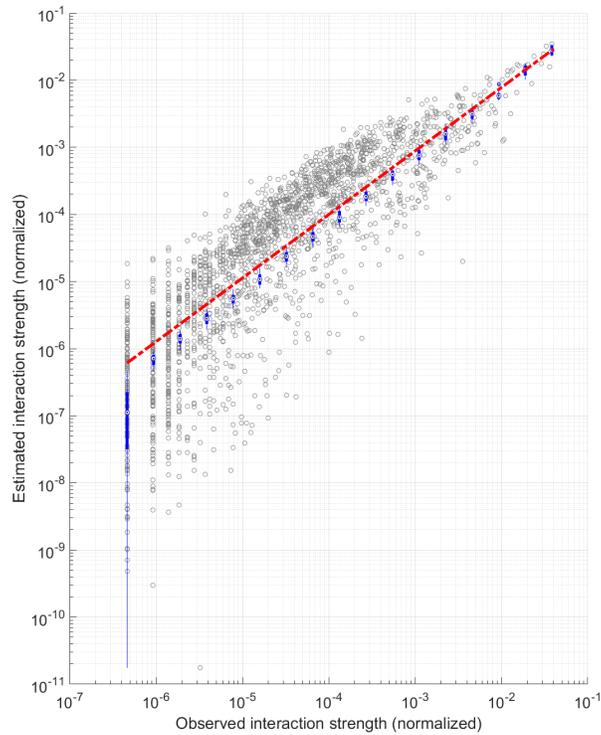

**Figure 7. The observed interaction strength versus the estimated ones from the adopted gravity model with $\beta = 0.8$** Note that the β value is taken from the probability density function of radius of gyration when $r_g > 10\ km$. The red dash line indicates strong linear correlation between the estimated and observed interaction strength with $R^2 = 0.89$ and $p-value < 0.01$.

Since we used different physical movement ranges of users in delineating the urban boundaries, the description length $L(M)$ appeared to get larger when longer displacements were used, which corresponded to less and larger geographically cohesive, non-overlapping urban areas. We believed that different movement ranges of users changed the weights of this graph and affected the interaction strength between two fishnet cells. As the employed gravity model suggested that the interaction strength between two fishnet cells decreases as the geographical distance increases, the longer displacements were used, the larger $L(M)$ were produced from the map equation algorithm.

## 5. Conclusion and Discussion

This paper developed a novel method for delineating non-administrative anthropogaphic urban boundaries by constructing a mobility network of Twitter user spatial interactions. In contrast to administrative urban boundaries, our "bottom-up" approach constructed a virtual fishnet over the islands of Great Britain. Twitter users' movements were used to establish a connectivity network of the fishnet cells. We applied the map equation algorithm to partition the network and associated geographical regions. The strongly connected communities within the network space yielded geographically cohesive, non-overlapping urban areas that provided a clear delineation of the urban boundaries in Great Britain. By performing a statistical analysis of the distribution of collective Twitter user displacements, we found multi-scale and multi-modal urban movements that were divided into several distance ranges starting from short intra-city to inter-city movements with clear destination points. Identifying the connected regions at each of these distance ranges revealed hierarchical boundaries of the urban space in Great Britain.

Using Twitter users' mobility to delineate non-administrative anthropographic boundaries enables urban representation at different mobility ranges inferred objectively from individual-based collective mobility patterns. Such urban boundaries capture physical commutes to reflect human spatial interactions in a geographical space. Importantly, as the depicted urban boundaries exhibit a strong instance of spatial proximity, we further employed a gravity model to connect human mobility research to understand the distance decay effects in shaping the delineated urban boundaries. This well-fitted gravity model explains how geographical distances found in the mobility patterns affect the interaction strength among different non-administrative anthropographic urban areas, gaining new insights into the interactions between human activities and urban geographic space.

It is worth noting that constructing a mobility network of spatial interactions using geo-located tweets has some potential limitations. First, the geo-located Twitter data do not represent the entire population. As the demographic information of Twitter users cannot be precisely identified, the results of delineated urban boundaries may not reflect a complete real-world image based on actual human movements. Related studies have examined this representation limitation (Steiger *et al.* 2015, Huang and Wong 2016, Luo *et al.* 2016), which emphasized the importance of carefully interpreting findings based on appropriate contexts and questions as done by our research. Second, regarding the spatial sparseness of geo-located Twitter data (Steiger *et al.* 2015), the urban regions that do not have any, or limited, Twitter coverage could be missed. To investigate whether this limited the ability to capture the connections made through Twitter user movements between urban regions, we visualized the flows of Twitter user movements using the method suggested by Rae (2009) (see Supplement Materials section 1). The outcome showed that Twitter users' movements in our research connected most urban areas in Great Britain and clearly captured long and short distance movements, which was essential for investigating the connection strength between urban regions. Third, since the collective radius of gyration was used to determine the fishnet cell size, we examined the temporal stability of this parameter. The probability distributions of the radius of gyration for Twitter users in Great Britain are verified to be consistent across different monthly time spans (see Supplement Materials section 2), which indicates the desirable stability.

As geo-located tweets are openly accessible, our method can be applied to other countries and regions. A major underpinning of the method is that geographical distance plays an important role in affecting human mobility patterns and the strength of human spatial interactions across urban space. The method can be considered to assist in understanding human spatial interactions from the mobility perspective, which is applicable based on detailed geo-located Twitter data in many countries, as well as future mobility datasets with location information of individuals and large spatial coverage.

*Acknowledgements*

We would like to thank the three anonymous reviewers for their constructive comments that helped improve the paper. We are grateful for insightful inputs to the manuscript received from Austin Davis and Ben Liebersohn in the CyberInfrastructure and Geospatial Information Laboratory at the University of Illinois at Urbana-Champaign. This material is based in part upon work supported by the U.S. National Science Foundation (NSF) under grant numbers: 1047916 and 1443080. Any opinions, findings, and conclusions or recommendations expressed in this material are those of the authors and do not necessarily reflect the views of NSF. The computational work used the ROGER supercomputer, which is supported by NSF under grant number: 1429699.

# Supplement Materials

## 1. Visualization of movement flux of Twitter users across Great Britain

In this study, our method relied on extracting Twitter user displacements/movements to build the mobility network of Twitter user spatial interactions across the Great Britain. To investigate the spatial coverage of the extracted Twitter user movements, we employed the visual-analytics method developed in (Rae 2009), where we plotted a flow map of the Twitter user displacements at the national level in Figure 1. In this figure, each yellow line represented one displacement/movement from a particular Twitter user. To achieve better visualization effects, this figure did not simply plot all the movements at once but highlighted those urban areas based on the density of movement flux. This visualization showed that the extracted Twitter user movements connected most urban areas in Great Britain and clearly exhibited long and short distance movements for connecting urban regions at different spatial scale, which was essential for building the mobility network.

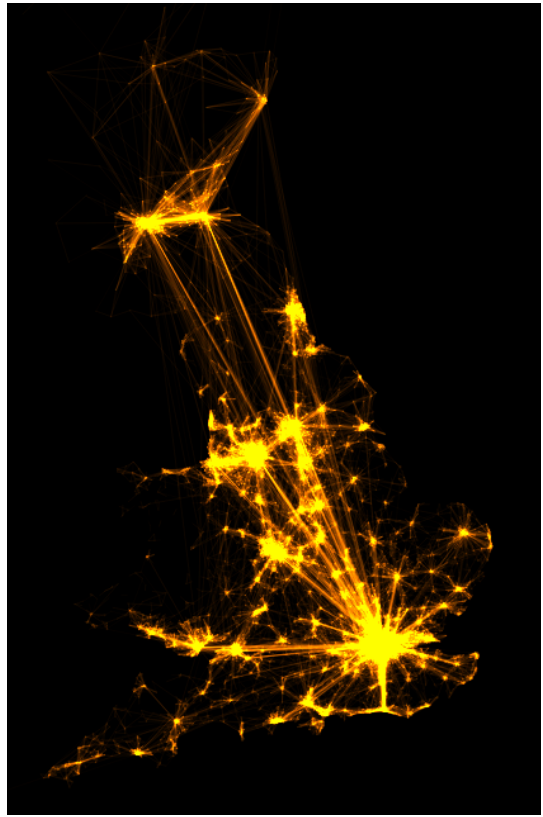

**Figure 1**: Visualization of the Twitter user movement flows across the Great Britain

## 2. Radius of gyration of Twitter users in Great Britain

The measurement of collective radius of gyration for individual Twitter users in the Great Britain was important for choosing the cell-size of the virtual fishnet. As the collected geo-located Twitter data in this study was from June 1st to December 31st, 2014, to investigate whether there were temporal fluctuations that would affect the consistency of such a measurement, we summarized the probability distribution of the radius of gyration for Twitter users in the Great Britain with a monthly interval. Figure 2 showed that the probability distributions of the radius of gyration for Twitter users were consistent throughout the 7-month time span, which indicated the stability of using such measurements in this study. Note that in these calculations, we did not apply the criteria to filter out "tourists" as we had suggested in this study.

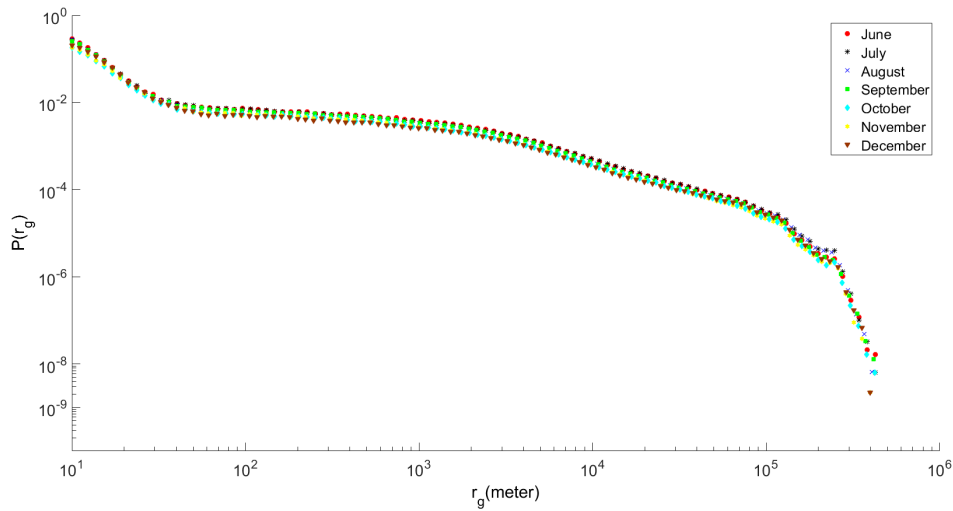

**Figure 2**: The probability distribution of the radius of gyration for Twitter users in each month

## 3. Mobility network based on ward divisions of the Great Britain

We imposed a virtual fishnet to partitioning the geographical space in the Great Britain. Such partition does not consider the underlying population information in each fishnet cell, therefore, it is worth to compare an alternative approach that does consider population information when partitioning the geographical space. In this study, we also carried out an experiment to partition the geographical space using a ward division of the Great Britain, which is the finest/smallest administrative boundary at the national level. This choice of using finest administrative boundary was also considered to minimize the conflict that administrative boundaries may not reflect natural human spatial interactions across space.

The strongly connected communities also yielded geographically cohesive, non-overlapping urban areas shown in Figure 3. The delineated urban boundaries were visually similar to the ones derived from using a fishnet approach. In particular, in the greater London region, the separate geographic areas that include Heathrow Airport exhibited more connectivity to central London than its surrounding areas (light orange region in Figure 3 (c)). However, as ward division is geographically continuous, it is problematic in aggregating regions that do not have Twitter coverage into certain clusters. Aggregating Twitter user movements at the ward level imposed more apparent concerns of the mismatch of the overall population, as long as there was one Twitter user movement fell into the polygonal unit, the entire unit would be considered in the mobility network. In this case, less populated areas were overly represented and connected into large areas, in particular, the delineation in Scotland (green region in Figure 3 (b)).

## 4. Mobility network with different settings

We chose a fishnet with 10 km cell-size to partition the geographical space of the Great Britain based on the statistical analysis of the probability distribution of the radius of gyration of Twitter users. Nevertheless, we also carried out an additional experiment by arbitrarily setting the fishnet cell-size to 5 km. The fishnet with smaller cell-size (i.e., 5 km) produced more and smaller strongly-connected communities within the network space as shown in Figure 4 (left). It suggested that the spatial resolution of the fishnet cells does affect the outcome from the community detection method employed in this study, where fishnet with smaller cell-size leads to more discrete and locally connected (i.e., smaller) clusters of urban areas. Such an effect can be explained by the probability distributions of the radius of gyrations of individual Twitter users. The probability of distance that deviates from a user's center location decays with a stretched-exponential function from [50m, 10km], which means the movements from Twitter users with smaller spatial coverage dominate the strength in connecting neighboring urban regions. Finally, we also illustrated that constructing the mobility network as an undirected graph did not lead to any meaningful delineation of the urban boundaries that reflect Twitter user spatial interactions, which is shown in Figure 4 (right).

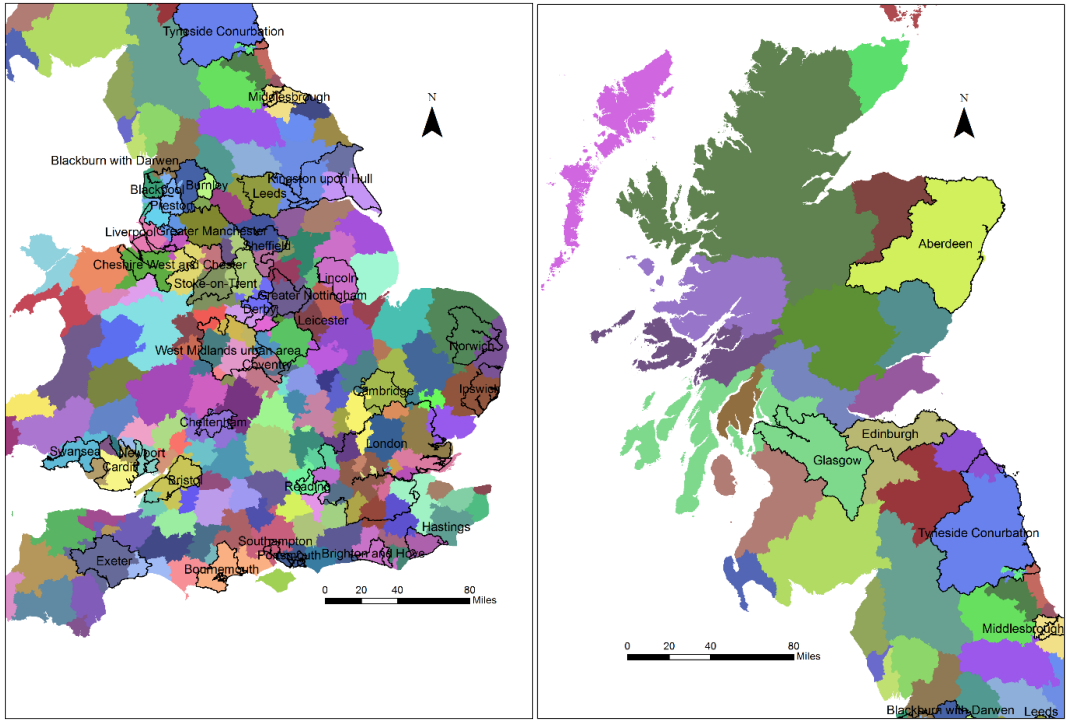
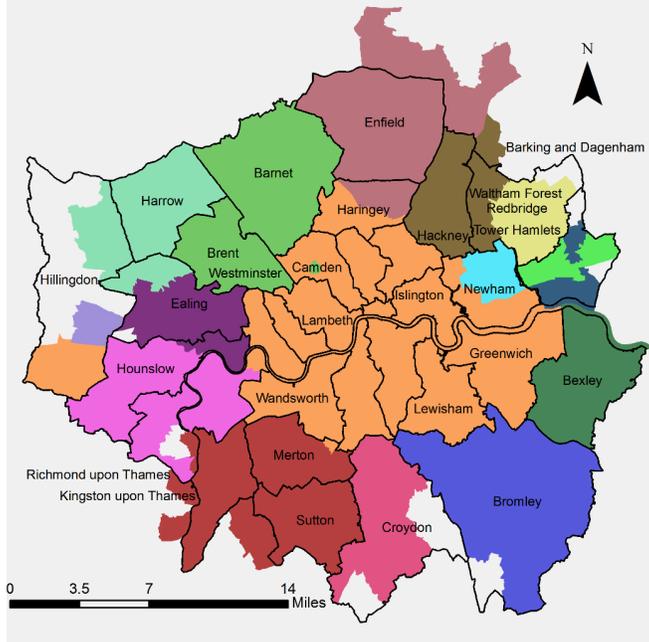

**Figure 3**: The community structure from collective Twitter user displacements based on the ward division in the Great Britain. (a) The delineation for England and Wales (b) The delineation for Scotland (c) The delineation for the greater London region

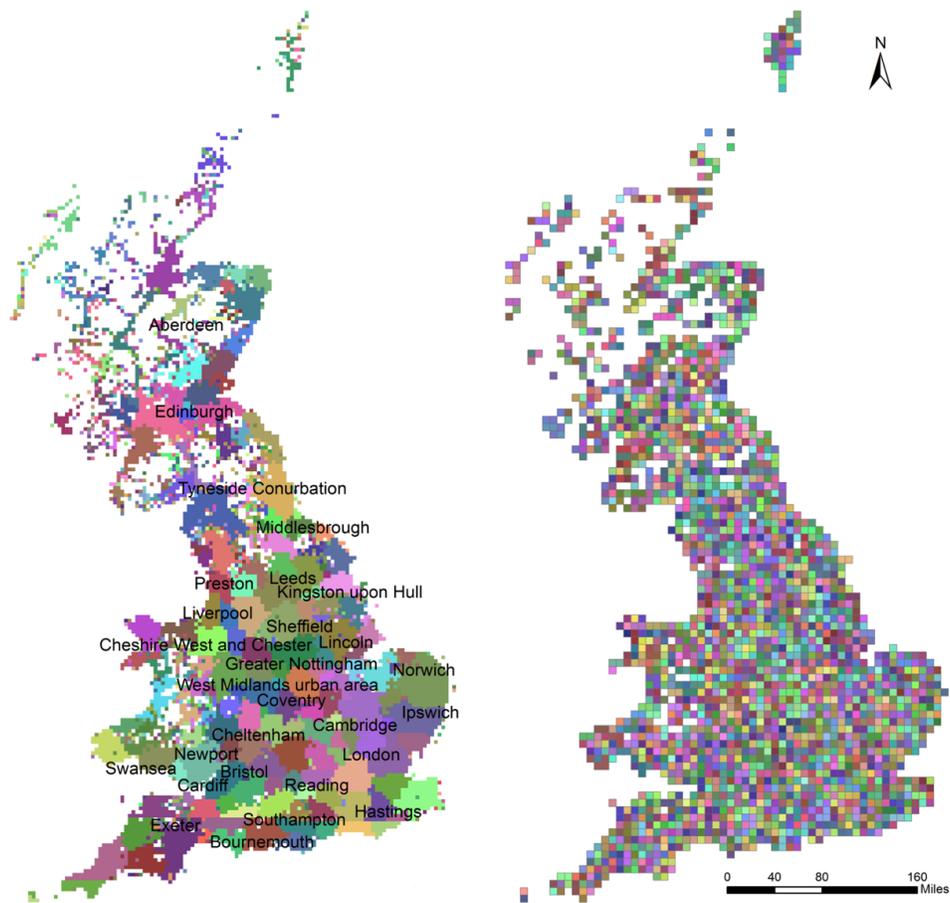

**Figure 4**: The community structure from collective Twitter user displacements with fishnet cell-size set to 5 km (left); and the community structure when the mobility network configured as undirected graph with fishnet cell-size of 10 km (right).